\documentclass[12pt]{article}

\newcommand{\be}{\begin{equation}}
\newcommand{\ee}{\end{equation}}
\newcommand{\ben}{\begin{eqnarray}}
\newcommand{\een}{\end{eqnarray}}
\newcommand{\n}{\label}

\def\AJL{{Ap. J. Lett.} }

\def\GRG{{Gen. Rel. Grav.} }

\def\JHEP{{JHEP} }

\def\NP{{Nucl. Phys.} }

\def\PR{{Phys. Rev.} }
\def\PRL{{Phys. Rev. Lett.} }

\begin{document}
\author{\large Juan M. Aguirregabiria$^1$\footnote{juanmari.aguirregabiria@ehu.es}\addtocounter{footnote}{2}\,, Luis P. Chimento$^2$\footnote{chimento@df.uba.ar}\addtocounter{footnote}{-2},  and Ruth Lazkoz$^1$\footnote{ruth.lazkoz@ehu.es}\\
{\it  $^1$ \small Fisika Teorikoa,  Zientzia eta Teknologia Fakultatea, }\\
{\it \small Euskal Herriko Unibertsitatea,}\\
{\it \small  644 Posta Kutxatila, 48080 Bilbao, Spain}\\{\it $^2$ \small Dpto. de F\'\i sica, Facultad de Ciencias Exactas y Naturales, }
\\{\it \small Universidad de Buenos Aires, Ciudad Universitaria,}\\
{\it \small  Pabell\'on I, 1428 Buenos Aires, Argentina}}
\title{Quintessence as k-essence}
\date{\today}
\maketitle
\begin{abstract}
Quintessence and k-essence have been proposed as candidates for the dark
energy component of the universe that would be responsible of the
currently observed accelerated expansion. In this paper we investigate
the degree of resemblance between those two theoretical setups, and find that every
quintessence model can be viewed as a k-essence model generated by a
kinetic linear function. In addition, we show the true effects of k-essence
begin at second order in the expansion of the kinetic function in
powers of the kinetic energy.

\end{abstract}
\titlepage
\section{Introduction}
Observations indicate that type Ia high redshift supernovae (SNIa) are
dimmer than expected \cite{supernovae}, and the mainstream
interpretation of this result is that the universe is currently
undergoing accelerated expansion driven by dark energy with negative
pressure. Further observations, like those of the Cosmic Microwave
Background (CMB) or Large Scale structures (LSS), suggest that two
thirds of the energy density of the universe correspond to dark energy.
Among several others, scalar field models have been proposed as candidates
for dark energy, and have therefore received significant attention.
Simplicity and economy has made researchers focus mainly on single field
cases, which fall into two classes: quintessence models (see
\cite{q_ess} for early papers) and k-essence models \cite{k_ess,k_ess2,cop,luis,series}
(the precursor of the concept of k-essence was k-inflation
\cite{k_inf}). The difference between those two setups is that k-essence
cosmologies, unlike quintessence ones, are derived form Lagrangians with
non-canonical kinematic terms. More specifically, given that the
equations of motion in all classical theories seem to be of second
order, the non-canonical terms considered in the Lagrangian will only be
combinations of the square of the gradient of the scalar field. 
 
Any suitable quintessence or k-essence model should provide a
satisfactory explanation to the cosmic coincidence problem (why the dark
energy component universe dominates only recently over the dark matter
one). One can devise situations with scalar fields with potentials that
go to zero asymptotically. These can have cosmologically interesting
properties, including ``tracking'' behavior that makes the current
energy density largely independent of the initial conditions, but
unfortunately the era in which the scalar field begins to dominate can
only be set by fine-tuning the parameters in the theory. A possible
remedy is to consider a dissipative matter component interacting with
dark energy \cite{pavon}. However, in k-essence models the solution seems not to
require the consideration of dissipation. Even for potentials that are not shallow, the nonlinear kinetic terms lead to dynamical attractor
behavior that permits the avoidance of the cosmic coincidence problem.
 
 Quintessence cosmologies have been exhaustively tested using  CMB and SNIa data mainly. This has resulted in constraints on the allowed shape of the quintessence potential.  In \cite{amenquer} it was found that Wilkinson Microwave Anisotropy Probe (WMAP) data alone constrain the equation of state of tracking dark energy to be $p/\rho\le-0.67$ (in compatibility with the bound given in \cite{mel}); this result implies for an inverse law potential an exponent smaller than $0.99$. Other works devoted to the same issue are \cite{coras} and \cite{balbi}, although they use the valuable but less refined BOOMERANG data set.
 
 As far as k-essence models are concerned,  the works which deal with observations and k-essence
from a general perspective are just a few \cite{Barger,eric}. In Ref. \cite{Barger} it is suggested that supernovae data alone
would not be able to distinguish between k-essence and quintessence.
Besides, in \cite{luis} it was discussed the correspondence between 
quintessence governed by a  exponential potential and  k-essence with a linear 
kinetic function $F$  driven by an inverse square potential. In that reference
it was imposed that the geometry generated by quintessence and  k-essence
be the same  (identical scale factor) together with the same requirement on the potential (specifically, that the potentials driving  quintessence and  k-essence be equal as function of cosmological time). These requirements lead to different but non independent fields for
quintessence and  k-essence.

In this paper we contribute to gaining more insight on the degree of
resemblance between quintessence and k-essence by extending the results
presented in \cite{luis} to quintessence driven by an arbitrary potential. First, in Section 2 we consider the case of a Friedman-Robertson-Walker geometry and homogeneous fields, and find which is the structure of the
kinetic function of the k-essence models which can be viewed as kinematically equivalent quintessence models, that is, as having the same geometry. This is done by imposing the validity of this equivalence for whichever quintessence and k-essence field, which means neither the quintessence nor the k-essence field depend on their derivative.  Then, in Section 3, starting from the knowledge gained in the simple homogeneous case, we study the situation for arbitrary spacetimes and inhomogeneous fields, and demonstrate that if the kinetic function has the same structure as in the earlier case then
the identification follows as well. Finally, in Section 4 we summarize our main results.
Our findings suggest that the debate of whether to opt for quintessence  or k-essence should rather be reformulated in terms of which is the most convenient type of k-essence. 

\section{Identification arising from geometry}

A possible way to compare quintessence and k-essence is through observations. As discussed in  \cite{Barger}, in order to fit the supernova data with a given quintessence or k-essence model, a choice of a
model-independent fitting function for the apparent magnitude $m(z)$
must be done. It turns out that the fitting function with the best fit
is derived using an expansion of the equation of state parameter $w(z)$
in powers of $z$, i.e., only kinematical aspects (the geometry) of the
problem are taken into account, and the outcome is an ambiguity that
makes it impossible two distinguish between the two theories.
Nevertheless, in \cite{Barger} the remark is made that since the speed
of sound of k-essence is not unity as in quintessence models perhaps an
analysis using CMB data would be able to detect some signal of
k-essence. The prospect of some success rests on the fact that in such
case dynamical aspects (the potential) would also be accounted for. Interestingly,
imposing  the dynamical condition that the 
quintessence and k-essence potentials be identical as in \cite{luis} does not remove the  ambiguity.

For all these reasons, we address the same problem from a more intrinsic
point of view. We first establish that for any quintessence model there
is a k-essence model which is kinematically equivalent to
the former, i.e., they share the same geometry and the same potential as
a function of the cosmological time. Note that our argument is different
from that in \cite{cop}, where the objective was to write any k-essence
model like a quintessence one. 

Let us restrict ourselves for the time being  to the cosmological setting corresponding
to a flat universe described by the Friedman-Robertson-Walker (FRW) metric.
The equations of motion for the gravitational field
 $g_{\mu\nu}$ in a universe with metric $ds^2=-dt^2+a^2(t)(dx^2+dy^2+dz^2)$ filled with an homogeneous  quintessence field (q-field) $\varphi$
minimally coupled to gravity as derived from the action
\begin{equation}
	S=-\int dx^4\sqrt{-g}\left(\frac{R}{2}+\frac{1}{2}\varphi_{,\mu}\varphi^{,\mu}+U(\varphi)\right),\label{qui}
\end{equation}
are the Einstein equations below:
\begin{eqnarray}
&&3H^2=\frac{1}{2}\dot\varphi^2+U(\varphi)\n{2},\\
&&\dot H=-\frac{1}{2}\dot\varphi^2 \n{3}.
\end{eqnarray}
In turn these equations imply
\begin{equation}
\n{kg}
\ddot{\varphi}+3H\dot\varphi+\frac{\partial U}{\partial  \varphi}=0,
\end{equation}
which is the Klein-Gordon equation for the scalar field $\varphi$.

In contrast, a k-field $\phi$  minimally coupled to gravity is defined by the action \cite{thesis}
\begin{equation}
S=-\int dx^4\sqrt{-g}\left(\frac{R}{2}+
{\cal L}_k(\phi,X)\right),
\end{equation}
where ${\cal L}_k(\phi,X)$ is an arbitrary function of $\phi$ and of the kinetic term $X=-\dot\phi^2$.
For this field the Einstein equations become (see again \cite{thesis})
\begin{eqnarray}
&&3H^2={\cal L}_k -2 X \frac{\partial{\cal L}_k}{\partial X}\n{6},
\\
&&\dot H=X\frac{\partial{\cal L}_k}{\partial X}\n{7},
\end{eqnarray}
while the k-field equation is

\be
\n{kgk}
\left[\frac{\partial{\cal L}_k}{\partial X}+2X
\frac{\partial^2{\cal L}_k}{\partial X^2}\right]\ddot\phi+
3H\frac{\partial{\cal L}_k}{\partial X}\dot\phi+
\frac{1}{2}\frac{\partial}{\partial\phi}\left[
{\cal L}_k-2X\frac{\partial{\cal L}_k}{\partial X}\right]=0.
\ee

Let us look for the conditions under which quintessence and k-essence
lead to the same geometry, i.e., the same scale factor. From Eqs. (\ref{2})-(\ref{3}) and (\ref{6})-(\ref{7}) the first necessary condition is
\begin{equation}
3H^2+\dot H=U(\varphi)={\cal L}_k(\phi,X) - X \frac{\partial{\cal L}_k(\phi,X)}{\partial X}\label{cond}.
\end{equation}
Under the assumption that $\varphi$ is an arbitrary function of $\phi$ and $X$, we now rewrite Eq. (\ref{kg}) and then demand the result is consistent with Eq. (\ref{kgk}) in the sense it does not lead to further conditions of the fields.

Combining Eqs. (\ref{3}) and (\ref{7}) one gets
\be
\dot\varphi=\sqrt{-2X\frac{\partial{\cal L}_k}{\partial X}}.
\n{dotdot}
\ee
Upon differentiation of the latter one obtains  an expression for $\ddot\varphi$
which after substitution in Eq. (4) and upon using condition (\ref{cond}) leads to
\be
\n{kg'}
\left[\frac{\partial{\cal L}_k}{\partial X}+X
\frac{\partial^2{\cal L}_k}{\partial X^2}\right]\ddot\phi+
3H\frac{\partial{\cal L}_k}{\partial X}\dot\phi+
\frac{1}{2}\frac{\partial}{\partial\phi}\left[
{\cal L}_k-2X\frac{\partial{\cal L}_k}{\partial X}\right]=0.
\ee
Consistency between Eqs. (\ref{kgk}) and (\ref{kg'}) requires that
\be
\frac{\partial^2{\cal L}_k}{\partial X^2}=0.
\ee
Hence, ${\cal L}_k$ must be of the form 
\be 
\n{ourlag}
{\cal L}_k=V(\phi) X+K(\phi),
\ee
with $K$ and $V$ arbitrary functions of the k field $\phi$ and
\be
\n{U=V}
U(\varphi)=K(\phi),
\ee 
after using  Eq. (\ref{cond}).
The relation between the $U$ and $K$ must be understood in the sense that they are the same when written as functions of cosmological time, but different when  written as functions of the individual fields. So,  whenever one of the fields is known,  Eq. (\ref{U=V}) fixes the other univocally. In addition, Eqs. (\ref{dotdot}) and (\ref{ourlag}) give  the following relationship between both 
fields
\begin{equation}
	\varphi=\int \sqrt{2V}d\phi\label{philink}.
\end{equation}
Conditions (\ref{U=V}) and (\ref{philink}) are 
necessary and sufficient for the kinematical 
equivalence of FRW quintessence and k-essence cosmologies.

Using Eqs. (\ref{ourlag})-(\ref{philink}) in Eq. (\ref{kg'}), we see that Eq. (\ref{kg}) reduces now to 
the k-field equation (\ref{kgk}) which now looks as
\be
\n{kgf}
\ddot\phi+3H\dot\phi+\frac{1}{2V}\left[\frac{dK}{d\phi}+\frac{dV}{d\phi}\,\,\dot\phi^2\right]=0.
\ee
Finally, it can be seen that the Lagrangians of quintessence and k-essence
\be
\n{lq}
{\cal L}_q=-\frac{1}{2}\dot\varphi^2+U(\varphi),
\ee
\be
\n{lk}
{\cal L}_k=-V(\phi)\dot\phi^2+K(\phi),
\ee
map into each other under transformations (\ref{U=V})-(\ref{philink}). These are the only transformations which preserve the order of the field equations and make those Lagrangians coincide. 

Since the k-essence theoretical setup generated by the Lagrangian
 (\ref{ourlag}) requires knowing the two functions $K$ and $V$ to control the k-field through the field equation  (\ref{kgf}), we can restrict the model by imposing that the  k-essence Lagrangian be factorizable (as usual). That means we will take ${\cal L}=V(\phi)F(X)$, where $V(\phi)$ is the potential governing the k-essence and $F=X+1$ is the kinetic function, which depends on the kinetic energy $X$ solely, so $K=V$. Note that the latter restriction does not alter the relation between quintessence and k-essence fields (\ref{philink}). Thus, the  k-field (\ref{kgf}) equation gets simplified to
\be
\n{kgf'}
\ddot\phi+3H\dot\phi+\frac{1+\dot\phi^2}{2V}\frac{dV}{d\phi}\,=0.
\ee

\begin{table}[h!]
\caption[crit]{ Some k-essence potentials  and their quintessence correspondence.}\label{crit}
\begin{center}
\begin{tabular}{cc}\\
\hline
\hline
\vspace{-0.3cm}
\\
  $V(\phi)$&$U(\varphi)$\\
\hline\vspace{-0.3cm}\\
$\lambda\,e^{{-\phi}}$&$\displaystyle\frac{\varphi^2}{8}$\\\vspace{0.2cm}
$\displaystyle\frac{\lambda}{1+\phi^2}$
         &$\displaystyle{\lambda }\,
 {\rm sech} ^2\bigg(\displaystyle\frac{\varphi}
        {{\sqrt{2\lambda}}
         }\bigg)$\\
         \vspace{0.2cm}
$\displaystyle\frac{\lambda}{\phi^{2}}$&${\lambda }\,{\rm{exp}\left({-{{\displaystyle\sqrt{\frac{2}{\lambda}}}\varphi}}\right)}$\\\vspace{0.2cm}
$\displaystyle\frac{\lambda}{\phi^{n}}\quad(n\ne2)$&$\displaystyle\left[\frac{{\left(\left( 2 - n \right) \varphi\right)^2}}
  {
    {8\lambda^{2/n} }}\right]^{n/(n-2)}$\\
   \vspace{0.2cm} $\displaystyle\frac{\lambda }{{\left( 1 + \cosh \phi  \right) }^2}$&$\displaystyle\frac{{\left( \varphi^2 - 2\lambda  \right) }^2}{16{\lambda }}$\\ 
\hline
\hline
\end{tabular}
\end{center}
\label{tabla}
\end{table}

At this stage, it is worth moving on and  illustrating our findings. We begin by outlining in Table \ref{tabla} some possible
potentials for k-essence models, and the corresponding potential for the
quintessence counterpart. On the first row we have
the exponential potential, which was proposed as a potential for the
tachyon by Sen \cite{exp}. On the next three rows we have  other
potentials proposed for the tachyon also. The first of those potentials
\cite{fraction} becomes constant for small $\phi$ but goes like
$\phi^{-2}$ for large $\phi$. We see that the associated quintessence
potential has a simple trigonometric expression. On the third row we have the pure inverse square potential
\cite{power}, which leads to an exponential quintessence potential
as shown in \cite{luis}. On
the fourth row we have a power-law potential with a negative exponent
\cite{power}, which for $n<1$ leads to a power-law k-essence potential
also with a negative exponent (recall that observations restrict the
exponent of power-law quintessence potentials to be smaller than 0.99).
Finally, on the last row, we present the k-essence potential which leads to the famous
double-well Duffing potential.

The list of potentials one could consider is neverending, but there is the limitation of physical motivation on one hand and computational feasibility on the other. One could for instance, consider the  potential \footnote{A larger class of potentials containing this particular one was considered in \cite{chimjaku}.}
\begin{equation}
U(\varphi)\propto \left(\cosh 3\varphi+\frac{1}{\cosh 3\varphi}\right)
\end{equation}
for action (\ref{qui}), because it
leads to a class of cosmological models which under some particular initial conditions \cite{kam} are
conventional Chaplygin cosmologies~\footnote{Different flavors of Chaplygin cosmologies have gathered much attention due to the role they play as unified dark matter models \cite{genchap,chapobs}}.
Unfortunately, the expression obtained under application of (16) is not invertible so this case is of little use and we will not consider it further.

>From this equivalence perspective, one might also one to have a look at generalized \cite{genchap} and modified Chaplygin cosmologies \cite{luis}. The bad news is that they seem to be derivable only from Born-Infeld Lagrangians, which do not have a canonical kinetic term, and therefore these two classes of Chaplygin cosmologies cannot be classified as quintessence cosmologies. Indeed, they are k-essence cosmologies, and more specifically representatives of the so-called class of purely kinetic k-essence cosmologies (see \cite{novello} and the references therein). This being so, what can be said about their equivalence to quintessence models? This is an issue which, in fact, extends to a larger set of Lagrangians, i.e., to that of factorizable Lagrangians.

Such a label corresponds to the Lagrangians of the form ${\cal L}_k=V(\phi)F(X)$, with $V$ and  $F$ arbitrary functions of $\phi$ and $X$, and they are naturally motivated by string theory. Interestingly, k-essence models derived from factorizable Lagrangians in which $F(X)$ is a linear function of $X$ mimic 
the behavior of other models. Let us illustrate it for the case of the tachyon, which corresponds to $F=(1+X)^{1/2}$. For $\vert X\vert\ll 1$ one has $F\approx 1+X/2$ and it leads to
the late time asymptotic of the scale factor. In  \cite{series} sets of cosmologies with $F$ functions admitting a power
series expansion in the form $F(X)=F(0)+F'(0)X+......$ were considered.  At first order in $X$ such models  behave like those one would obtain from (\ref{ourlag})  and the quintessence effects will be more important than the k-essence ones. In contrast, 
effects strictly due to the actual k-essence nature of the model will begin to become non-negligible when the condition $\vert X\vert\ll 1$ breaks down.

\section{Covariant proof for arbitrary spacetimes}

In the last section, we have established the conditions for the
kinematical  equivalence between FRW quintessence and
k-essence cosmologies. In what follows we are going to use the insight gained in the previous section regarding the structure of ${\cal L}_k$ so as to  demonstrate equivalent results for an
arbitrary spacetime. 

Let us begin by imposing the condition that the geometry generated either by quintessence or k-essence be the same. Put another way, this means we are demanding the
 quintessence Einstein tensor $G_{\mu\nu}^{(q)}$ be the same as the k-essence one $G_{\mu\nu}^{(k)}$, thus
\be
\n{G}
G_{\mu\nu}^{(q)}=T_{\mu\nu}^{(q)}\equiv G_{\mu\nu}^{(k)}=T_{\mu\nu}^{(k)},
\ee
where on the one hand $T_{\mu\nu}^{(q)}$ is the stress-energy tensor of quintessence, 
\be
\n{tq}
T_{\mu\nu}^{(q)}=\varphi_{,\mu}\varphi_{,\nu}-g_{\mu\nu}\left(\frac{1}{2}\varphi_{,\sigma}\varphi^{,\sigma}
+U(\varphi)\right),
\ee
and on the other hand $T_{\mu\nu}^{(k)}$ is the stress-energy tensor of k-essence, 
\be
\n{tk}
T_{\mu\nu}^{(k)}=2\frac{\partial{\cal L}_k}{\partial X}\phi_{,\mu}\phi_{,\nu}-g_{\mu\nu}{\cal L}_k,
\ee
where $X=\phi_{,\mu}\phi^{,\mu}$.
 If we now rewrite the latter using (\ref{ourlag}),  and then compare with Eq. (\ref{tq}),
it  follows from  identity (\ref{G}) that the following two relations must hold:
\be
\n{uvg}
U(\varphi(x^{\mu}))=V(\phi(x^{\mu})),
\ee

\be
\n{fig}
\varphi_{,\mu}=\sqrt{2V(\phi(x^{\mu}))}\,\phi_{,\mu}.
\ee
Multiplying by $dx^{\mu}$
we get $d\varphi=\sqrt{2V(\phi)}\,d\phi$, and by integration we obtain the following prescription to relate the fields:
\be
\n{final}
\varphi(x^{\mu})=\int \sqrt{2V(\phi(x^{\mu}))}\,d\phi.
\ee
This generalizes the relation (\ref{philink}) obtained previously in a
more restrictive case
to situations in which the fields depend on both space and time
coordinates. 

In addition, taking into account that the energy density and the
pressure of the k-essence fluid are $ \rho={\cal L}_k-2(\partial{\cal L}_k/\partial X)$ and
$p=-{\cal L}_k$ respectively, one can see that the  sound speed  $c_s^2=(dp/dX)/(d\rho/dX)=1$ of a k-essence model with (\ref{ourlag})  coincides with the sound speed of the quintessence fluid, so this completes the proof. This result is in agreement
with Ref. \cite{eric} where it was shown that any scalar field action with a linear kinetic term has a speed of sound equal to one. 

\section{Conclusions}

Let us come to conclusions and discussion now. Quintessence and
k-essence  are not the only dark energy candidates
proposed so far, but they are very popular, particularly the former. At
this stage it is important to understand not only from the observational
point of view but also from a more fundamental one the degree of
resemblance of these two setups. In what regards observations, it has
already been discussed that supernovae data alone are unlikely to be
able to do such discrimination. In contrast, if one combines CMB and supernovae data some hint of non-equivalence could be obtained. In broad
terms this is due to the fact scalar perturbations of  quintessence and k-essence models 
do not follow the same rules (i.e., the corresponding theoretical frameworks are dynamically inequivalent). We think, however, that this topic has not
been addressed in the literature in sufficient depth, and we hope our work
contributes to enlighten it.

We have first demonstrated that any quintessence is contained into
k-essence frame with a linear kinetic function, and we have obtained the
prescription that gives the q-field in terms of the k-field
(this can be used to relate the potentials of the two models). Then we
have turned to the Einstein field equations for an arbitrary spacetime
and we have proved simply and neatly the theoretical frame of quintessence can be fully included into that of k-essence  by extending the previously obtained
relation among the homogeneous fields. Thus, each quintessence model is
kinematically equivalent to a k-essence model.

An interesting related result is that the true effects of k-essence
begin at second order in the expansion of the kinetic function in
powers of the kinetic energy.

Finally, coming back to the issue of observations, in the light of our results we can say that a combination of CMB and supernovae data is not going to tell us whether k-essence is preferable to quintessence, but rather what sort of k-essence is admissible 
(the one generated by a linear kinetic function or other alternative).

\section*{Acknowledgments}

Thanks to  A. D\'\i ez Tejedor for conversations. L.P.C. is partially funded by the University of Buenos Aires  under
project X224, and the Consejo Nacional de Investigaciones Cient\'{\i}ficas y
T\'ecnicas.  J.M.A. and R.L. are supported by the University of the Basque Country through research grant UPV00172.310-14456/2002, and
by the Spanish Ministry of Science and Education through research grant FIS2004-01626. 

\end{document}